\documentclass[10pt,prd,superscriptaddress,preprintnumbers,amsmath,amssymb,nofootinbib]{revtex4}
\usepackage{epsfig}  
\usepackage{graphicx}
\usepackage{hyperref}
\usepackage{color}
\usepackage{float}
\usepackage{amsfonts}
\usepackage{amsmath}
\usepackage{slashed}
\usepackage{soul}

\large

\newcommand{\ba}{\begin{array}}
\newcommand{\ea}{\end{array}}
\newcommand{\bea}{\begin{eqnarray}}
\newcommand{\eea}{\end{eqnarray}}

\newcommand{\be}{\begin{equation}}
\newcommand{\ee}{\end{equation}}

\def\Bbar{\overline{B}}

\def\cbar{\overline{c}}

\def\Dst{{D^*}}

\begin{document} 
\title{$D^*$ polarization as a probe to discriminate new physics in $\bar{B} \to D^* \tau \bar{\nu}$}

\author{Ashutosh Kumar Alok}
\email{akalok@iitj.ac.in}
\affiliation{Indian Institute of Technology Jodhpur, Jodhpur 342011, India}

\author{Dinesh Kumar}
\email{dinesh@phy.iitb.ac.in}
\affiliation{Indian Institute of Technology Bombay, Mumbai 400076, India}
\affiliation{University of Rajasthan, Jaipur 302004, India}

\author{Suman Kumbhakar}
\email{suman@phy.iitb.ac.in}
\affiliation{Indian Institute of Technology Bombay, Mumbai 400076, India}

\author{S Uma Sankar}
\email{uma@phy.iitb.ac.in}
\affiliation{Indian Institute of Technology Bombay, Mumbai 400076, India}

\date{\today} 

\preprint{}

\begin{abstract}
The confirmation of excess in $R_{D^*}$ at the LHCb is an indication of lepton flavor 
non-universality. Various different new physics operators and their coupling strengths, 
which provide a good fit to $R_D$, $R_{D^*}$ and $q^2$ spectra, were identfied previously.
In this work, we try to find angular observables in $\bar{B} \to D^* \tau \bar{\nu}$ which enable us to distinguish between these new physics operators. 
We find that the $D^*$ polarization fraction $f_L(q^2)$ is a 
good discriminant of scalar and tensor new physics operators. 
The change in $\langle f_L(q^2) \rangle$, induced by scalar and tensor operators, is about three times larger than the expected uncertainty in the upcoming Belle measurement. 
 
\end{abstract}

\maketitle

\section{Introduction}
The currently running LHC has not only provided new signatures of possible physics beyond the Standard Model (SM) but also confirmed some of the prevailing tensions in the SM. The most striking example of the confirmation of previously observed anomaly is the 3.5$\sigma$ deviation from the SM expectation of the ratio $R_{D^{*}}=\Gamma(B \to D^*\, \tau \bar{\nu})/\Gamma(B \to D^*\, l \bar{\nu})$ $(l=e,\,\mu)$\cite{Lees:2012xj,Lees:2013uzd,Adachi:2009qg,Bozek:2010xy,Aaij:2015yra,Huschle:2015rga}. This is an indication towards lepton flavor non universality, in disagreement with the SM predictions. The idea of lepton flavour non universality is further bolstered by the measurement that $R_K = \Gamma(B \to K \,\mu^+\,\mu^-)/\Gamma(B \to K\,e^+\,e^-) $ \cite{rk} and $R_{K^*} = \Gamma(B \to K^* \,\mu^+\,\mu^-)/\Gamma(B \to K^*\,e^+\,e^-) $ \cite{rkstar} are not equal to unity.

The four fermion interaction $b \to c\, \tau\, \bar{\nu}$, which induces the decays $B \to (D,\,D^*)\, \tau \bar{\nu}$, occurs at the tree level within the SM. Note that the situation in the case of $b \to s\, \mu^+\, \mu^-$ is quite different because this transition occurs only at one loop level in SM.
Relatively large new physics (NP) contributions are required to explain the anomaly in the measurement of $R_{D^{*}}$. Such large contributions are more likely to occur in NP models where the four fermion interaction occurs at tree level. However such models must also be consistent with the measurement of other observables which are in agreement with their SM predictions. As a result there are only a limited set of NP models which can explain the $R_{D^{*}}$ anomaly, see for example \cite{Fajfer:2012jt,Crivellin:2012ye,Celis:2012dk,Tanaka:2012nw,Dorsner:2013tla,Sakaki:2013bfa,Bhattacharya:2014wla,Alonso:2015sja,Calibbi:2015kma,Freytsis:2015qca,Hati:2015awg,Fajfer:2015ycq,Bauer:2015knc,Barbieri:2015yvd,Boucenna:2016wpr,Deshpand:2016cpw,Sahoo:2016pet,Wang:2016ggf,Alonso:2016oyd}. In particular, ref. \cite{Freytsis:2015qca} listed all the four fermion operators contributing to $B\rightarrow D^* \tau \bar{\nu}$ and derived the values of various NP couplings which satisfy the measurement of $R_D,R_{D^*}$ and the $q^2$ spectra.  

The next step is to discriminate between various NP operators which can explain the excess in 
$R_{D^{*}}$.  This can be achieved if we have a handle on various angular observables in 
$B \to D^*\, \tau \bar{\nu}$ similar to the ones we have in $B \to K^* \, \mu^+\,\mu^-$ decay. 
In the semileptonic decays of pseudoscalar mesons to vector mesons, it is possible to measure differential distributions with respect to three angles, besides $d\Gamma / dq^2$. These angles are usually defined in the vector meson rest frame. For the decay $B \rightarrow D^* \tau \, \bar{\nu}$, these angles are (a) $\theta_D$, the angle between $B$ and $D$ where the $D$ meson comes from $D^*$ decay, (b) $\theta_{\tau}$, the angle between $\tau$ and $B$ and (c) $\phi$, the angle between $D^*$ decay plane and the plane defined by the lepton momenta \cite{Alok:2010zd}.
A study of these angular distributions, in the case of $B \rightarrow K^* \mu^+\, \mu^-$, has revealed significant discrepancies between the measurements and the predictions of SM \cite{Aaij:2013iag}. Various authors have done theoretical analysis of similar angular distributions for $B \rightarrow D^* \tau \, \bar{\nu}$~\cite{Fajfer:2012vx,Sakaki:2012ft,Datta:2012qk,Biancofiore:2013ki,Duraisamy:2013kcw,Duraisamy:2014sna,Sakaki:2014sea,Bhattacharya:2015ida,Becirevic:2016hea,Alonso:2016gym,Ivanov:2016qtw,Ligeti:2016npd,Bardhan:2016uhr,Kim:2016yth,Dutta:2016eml,Bhattacharya:2016zcw}.

So far the $\tau$ lepton  has not been reconstructed in any of the experiments which measured $R_D$ and $R_{D^*}$\footnote{Recently the Belle collaboration reported their measurement of $\tau$ polarization in the $B \rightarrow D^* \tau \, \bar{\nu}$ decay \cite{Abdesselam:2016xqt}. Note that this measurement did not involve reconstruction of $\tau$.}. Therefore $\theta_{\tau}$ and $\phi$ have not been measured. Hence it is not possible to measure the full differential distribution with present data \footnote{In future, it may be possible 
to estimate the $\tau$ momentum by considering $\tau$ decays into multi-hadron final states.}.  However, it is possible to measure $\theta_D$ and hence determine the $D^*$ polarization fraction $f_L(q^2)$. In fact, the Belle Collaboration is in the process of making this 
measurement \cite{belle-ckm16}. We calculate  $f_L(q^2)$  for all the NP solutions which account for $R_{D^*}$ excess and show that it can discriminate against NP solutions with scalar and tensor operators. We also find that the forward-backward asymmetry, $A_{\rm FB}(q^2)$, has a discrimination capability similar to that of $f_L(q^2)$. However, measuring this quantity is  more difficult as it requires the reconstruction of the $\tau$ lepton.

\section{Disentangling various new physics contributions to $\bar{B} \to D^* \tau \bar{\nu}$}

First we summarize the results of ref.~\cite{Freytsis:2015qca}, which performed a fit of various NP models to the present  $R_{D^{*}}$  and $R_{D}$ data. These fits are also consistent with the $q^2$ distribution provided by BaBar \cite{Lees:2013uzd} and Belle \cite{Huschle:2015rga}.
The effective Hamiltonian for the quark level transition $b \to c\, \tau \, \bar{\nu}$ is given by 
\begin{equation}
H_{eff}= \frac{4 G_F}{\sqrt{2}} V_{cb}\left[O_{V_L} + \frac{\sqrt{2}}{4 G_F V_{cb}} \frac{1}{\Lambda^2} \left\lbrace \sum_i \left(C_i O_i +
 C^{'}_i O^{'}_i + C^{''}_i O^{''}_i \right) \right\rbrace \right],
\label{effH}
\end{equation}
where the scale $\Lambda$ is assumed to be 1 TeV. The Lorentz structures of the unprimed $\mathcal{O}_i$ and primed $\mathcal{O}^{'}_i$ and $\mathcal{O}^{''}_i$ operators are given in Table~\ref{operator}. For each primed operator, this table also lists the corresponding Fierz transformed unprimed operator.
 In the above analysis, the NP operators were considered either
one at a time or two similar operators (either $V \pm A$ or $S \pm P$) at a time. This
was done to obtain the strongest possible constraints on the coefficients of NP
operators from limited data. The values of coefficients of different NP
operators, which provide a good fit to the data, are given in Table~\ref{tab1}.

\begin{table}[h!]
\begin{center}
\begin{tabular}{|l|ccc|}
  \hline\hline
	& Operator & & Fierz identity\\
  \hline
$\mathcal{O}_{V_L}$   & $(\bar{c} \gamma_\mu P_L b)\,(\bar{\tau} \gamma^\mu P_L \nu)$ & & \\
$\mathcal{O}_{V_R}$   & $(\bar{c} \gamma_\mu P_R b)\,(\bar{\tau} \gamma^\mu P_L \nu)$ & & \\
  $\mathcal{O}_{S_R}$   & $(\bar{c} P_R b)\,(\bar{\tau} P_L \nu)$ & & \\
  $\mathcal{O}_{S_L}$   & $(\bar{c} P_L b)\,(\bar{\tau} P_L \nu)$ & &\\
  $\mathcal{O}_T$       & $(\bar{c}\sigma^{\mu\nu}P_L b)\,(\bar{\tau}\sigma_{\mu\nu}P_L \nu)$ & &\\[2pt]
  \hline
 $\mathcal{O}'_{V_L}$ &$(\bar{\tau} \gamma_\mu P_L b)\,(\bar{c} \gamma^\mu P_L \nu)$ &
   $\longleftrightarrow$ &$\mathcal{O}_{V_L}$\\
$\mathcal{O}'_{V_R}$  & $(\bar{\tau} \gamma_\mu P_R b)\,(\bar{c} \gamma^\mu P_L \nu)$ &
    $\longleftrightarrow$ & $-2\mathcal{O}_{S_R}$ \\
 $\mathcal{O}'_{S_R}$  & $(\bar{\tau} P_R b)\,(\bar{c} P_L \nu)$ &
    $\longleftrightarrow$ & $-\frac{1}{2}\mathcal{O}_{V_R}$ \\
  $\mathcal{O}'_{S_L}$  & $(\bar{\tau} P_L b)\,(\bar{c} P_L \nu)$ &
    $\longleftrightarrow$ & $-\frac{1}{2}\mathcal{O}_{S_L} - \frac{1}{8}\mathcal{O}_T$\\
$\mathcal{O}'_T$      & $(\bar{\tau}\sigma^{\mu\nu}P_L b)\,(\bar{c}\sigma_{\mu\nu}P_L \nu)$ &
    $\longleftrightarrow$ & $-6\mathcal{O}_{S_L} + \frac{1}{2}\mathcal{O}_T$ \\[2pt]
  \hline
 $\mathcal{O}''_{V_L}$ & $(\bar{\tau} \gamma_\mu P_L c^c)\,(\bar{b}^c \gamma^\mu P_L \nu)$ &
    $\longleftrightarrow$ & $-\mathcal{O}_{V_R}$  \\
  $\mathcal{O}''_{V_R}$ & $(\bar{\tau} \gamma_\mu P_R c^c)\,(\bar{b}^c \gamma^\mu P_L \nu)$ &
    $\longleftrightarrow$ & $-2\mathcal{O}_{S_R}$ \\
$\mathcal{O}''_{S_R}$ &$(\bar{\tau} P_R c^c)\,(\bar{b}^c P_L \nu)$ &
$\longleftrightarrow$ & $\frac{1}{2}\mathcal{O}_{V_L}$ \\
$\mathcal{O}''_{S_L}$ & $(\bar{\tau} P_L c^c)\,(\bar{b}^c P_L \nu)$ &
    $\longleftrightarrow$ & $-\frac{1}{2}\mathcal{O}_{S_L} + \frac{1}{8}\mathcal{O}_T$ \\
 $\mathcal{O}''_T$     & $(\bar{\tau}\sigma^{\mu\nu}P_L c^c)\,(\bar{b}^c\sigma_{\mu\nu}P_L \nu)$ &
    $\longleftrightarrow$ & $-6\mathcal{O}_{S_L} - \frac{1}{2}\mathcal{O}_T$  \\[2pt]
  \hline\hline
\end{tabular}
\caption{All possible four-fermion operators that can contribute to $\bar B \to
D^{(*)} \tau\bar{\nu}$.}
\label{operator}
\end{center}
\end{table}

\begin{table}[h!]
\begin{center}
\begin{tabular}{|c|l|l|}
\hline\hline
Coefficient(s)  &  Best fit value(s)  & $\langle f_L(q^2) \rangle$ \\
\hline
$C_{V_L}$  &  $0.18 \pm 0.04$& $0.46 \pm 0.04$  \\
$C_{V_L}$  &  $-2.88 \pm 0.04$ & $0.46 \pm 0.04$  \\
\hline
$C_T$  &  $0.52 \pm 0.02$, & $0.14 \pm 0.04$    \\
$C_T$  &  $-0.07 \pm 0.02$ , & $0.45 \pm 0.04$    \\
\hline
$C''_{S_L}$  &  $-0.46 \pm 0.09$  & $0.46 \pm 0.04$ \\
\hline
$(C_{S_R},\, C_{S_L})$  &  $(1.25,-1.02)$, & $0.41\pm 0.04$ \\
$(C_{S_R},\, C_{S_L})$  &  $(-2.84,3.08)$ & $0.76 \pm 0.04$ \\
\hline
$(C'_{V_R},\, C'_{V_L})$  &  $(-0.01,0.18)$ & $0.46 \pm 0.04$ \\
$(C'_{V_R},\, C'_{V_L})$  &  $(0.01,-2.88)$ & $0.46 \pm 0.04$ \\
\hline
$(C''_{S_R},\, C''_{S_L})$  &  $(0.35,-0.03)$ & $0.46\pm0.04$\\
$(C''_{S_R},\, C''_{S_L})$  &  $(0.96,2.41)$ & $0.26 \pm 0.04$ \\
$(C''_{S_R},\, C''_{S_L})$  &  $(-5.74, 0.03)$ & $0.46 \pm 0.04$ \\
$(C''_{S_R},\, C''_{S_L})$  & $(-6.34,-2.39)$ & $0.27 \pm 0.04$ \\
\hline\hline
\end{tabular}
\caption{Best fit values of the new physics operator coefficients which provide a good fit to the present experimental data in $b \to c\, \tau \, \bar{\nu}$ sector \cite{Freytsis:2015qca}. The values in the upper (lower) panel are obtained by considering one (two) new physics operator(s) at a time in the fit. For each case,
the corresponding predicion for $\langle f_L (q^2) \rangle $ is also listed. Note that for SM, $\langle f_L(q^2) \rangle$ is same as that of the $\mathcal{O}_{V_L}$.}
\end{center}
\label{tab1}
\end{table}

 In the fit with the operators $\mathcal{O}''_{S_R}$ and $\mathcal{O}''_{S_L}$, there are four allowed values for $C''_{S_R}$ and $C''_{S_L}$ couplings.  An attempt is made in ref.~\cite{Li:2016vvp} to distinguish between these solutions based on their predictions for the rate of 
the decay $B^-_c \rightarrow\tau^-\bar{\nu}$. They found that the two solutions 
$(C''_{S_R},\, C''_{S_L})=(0.96,2.41)$ and $(-6.34,-2.39)$ are excluded because the predicted leptonic partial decay width of $B_c$ meson is larger than its measured total decay width. This result can be understood by noting that $\mathcal{O}''_{S_R}$ operator is equivalent to $\mathcal{O}_{V_L}$ whereas $\mathcal{O}''_{S_L}$ operator is equivalent to linear combination of $\mathcal{O}_{S_L}$ and $\mathcal{O}_T$.
The two solutions with $(C''_{S_R},\, C''_{S_L})=(0.35,-0.03)$ and 
 $(-5.74,0.03)$ have essentially $\mathcal{O}_{V_L}$ Lorentz structure. Therefore their prediction for the pure leptonic decay of $B_c$ is subject to helicity suppression. For the other two cases, ruled out by  \cite{Li:2016vvp}, the coefficient of $\mathcal{O}''_{S_L}$ and hence $\mathcal{O}_{S_L}$ is quite large. For this operator there is no helicity suppression which leads to the prediction of very large decay width for $B^-_c \rightarrow\tau^-\bar{\nu}$. 

The angular distribution in $\theta_D$ is given by \cite{Duraisamy:2013kcw}
\begin{equation}
\frac{d^2\Gamma}{dq^2 d\cos{\theta_D}}  =  \frac{3}{4}\frac{d\Gamma}{dq^2}
\left[2f_L(q^2)\cos^2{\theta_D} +  \{ 1-f_L(q^2) \} \sin^2{\theta_D}\right], 
\label{d2G}
\end{equation}
where the $f_L(q^2)$ is defined to be
\begin{equation}
f_{L}(q^2)= \frac{A_L}{A_L + A_T}.
\end{equation}
The quantities $A_L$ and $A_T$ are defined in \cite{Alok:2010zd}. From these formulae we compute $f_L(q^2)$ and $\langle f_L (q^2) \rangle$ for the allowed NP couplings listed in Table~\ref{tab1} \cite{Freytsis:2015qca}. 

\begin{figure*}[h!] 
\centering
\begin{tabular}{ccc}
\includegraphics[width=35mm]{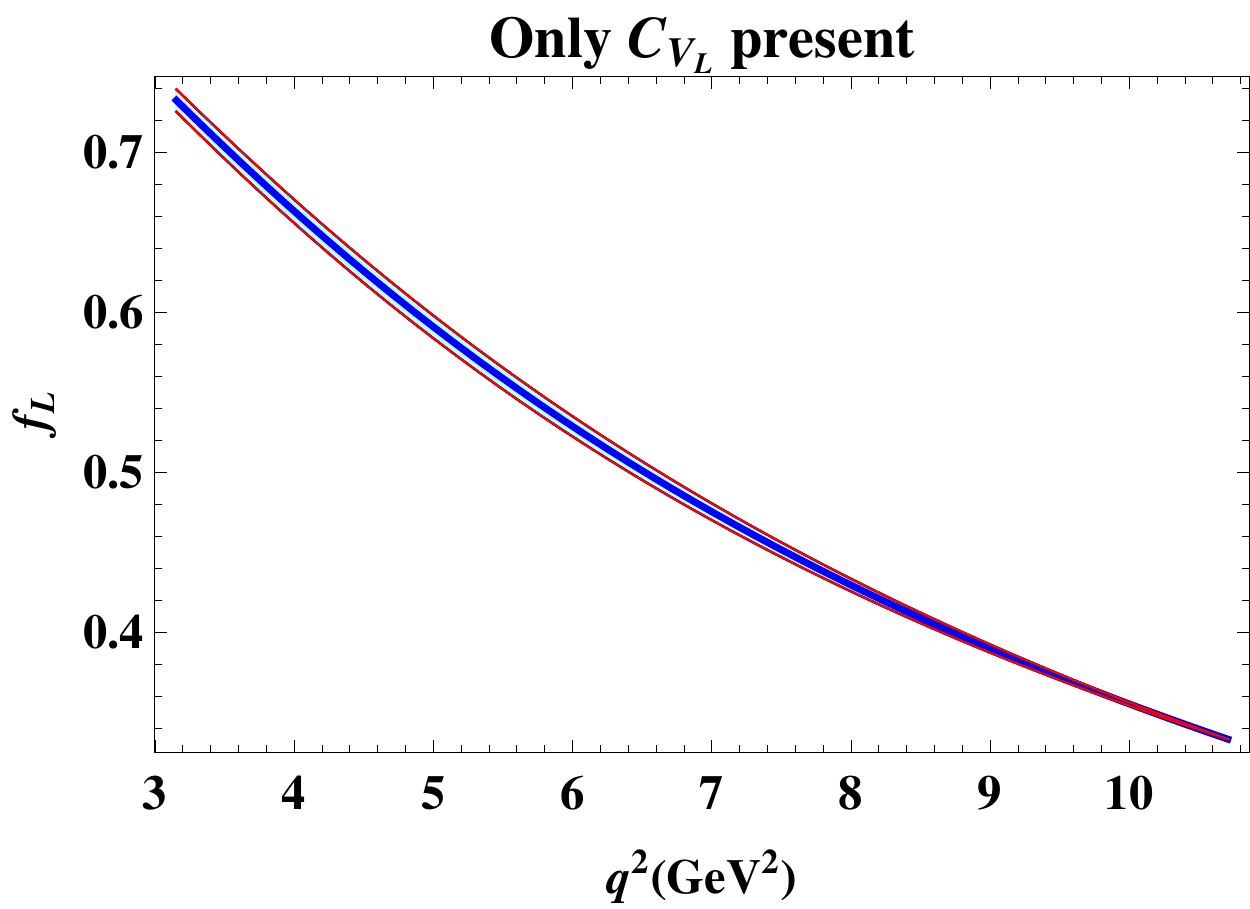}&
\includegraphics[width=35mm]{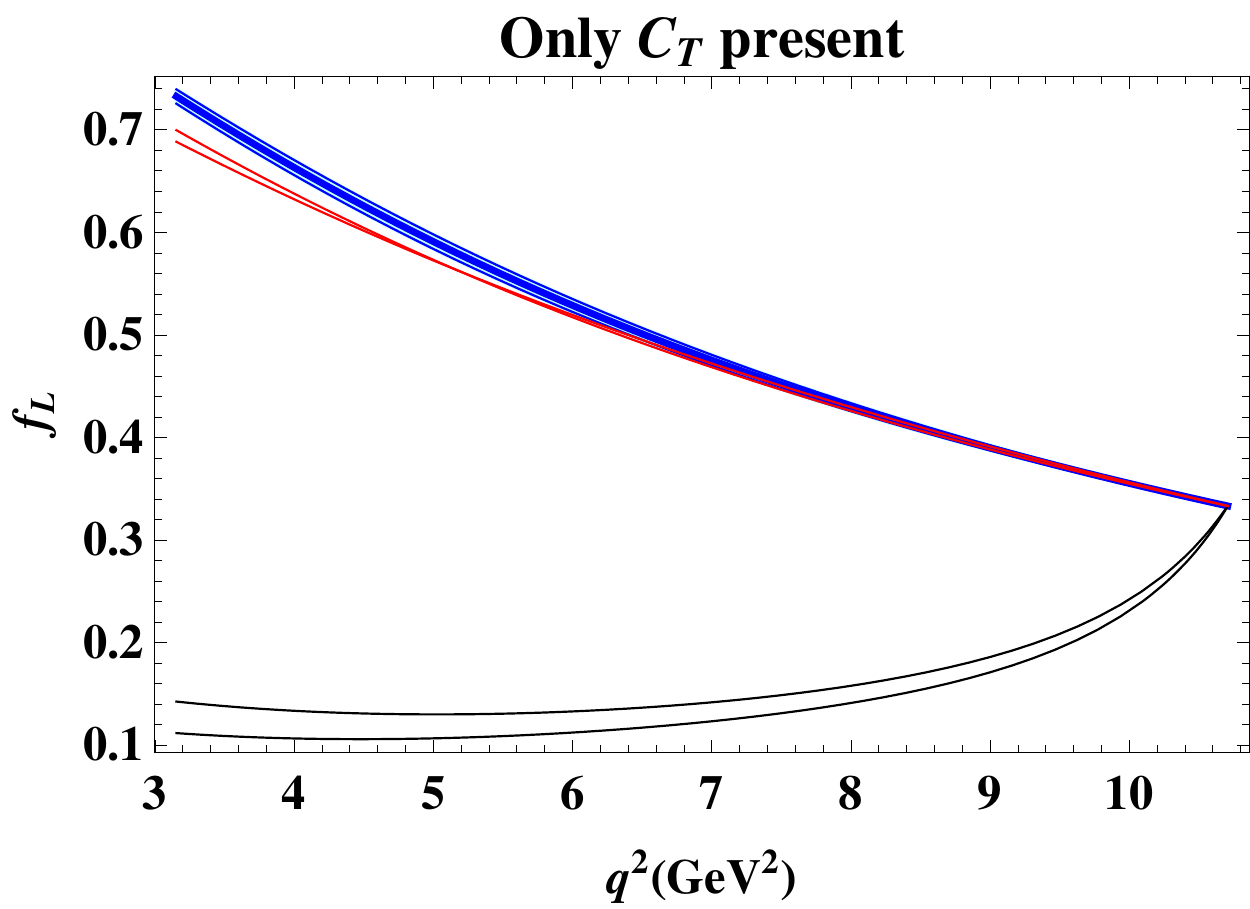}&
\includegraphics[width=35mm]{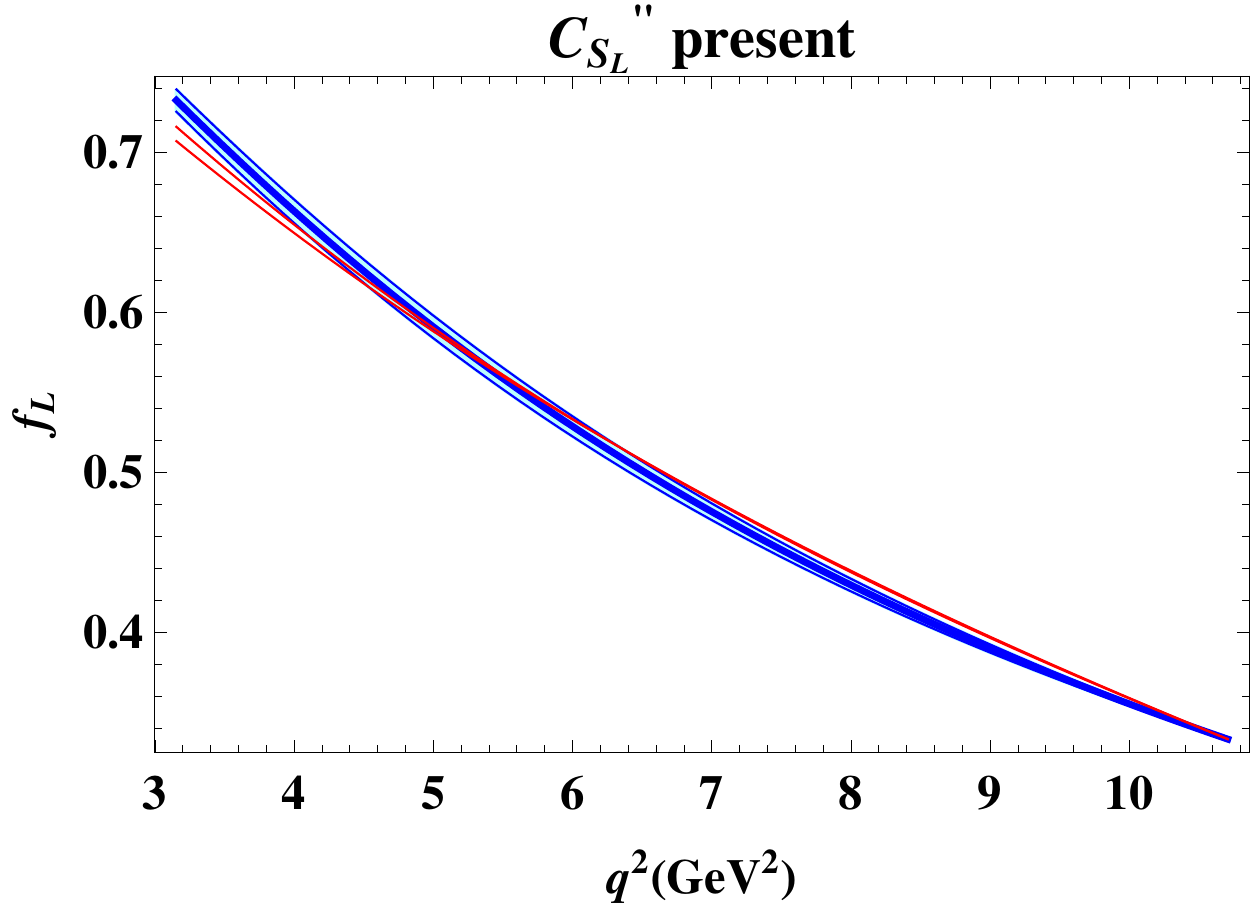}\\
\includegraphics[width=35mm]{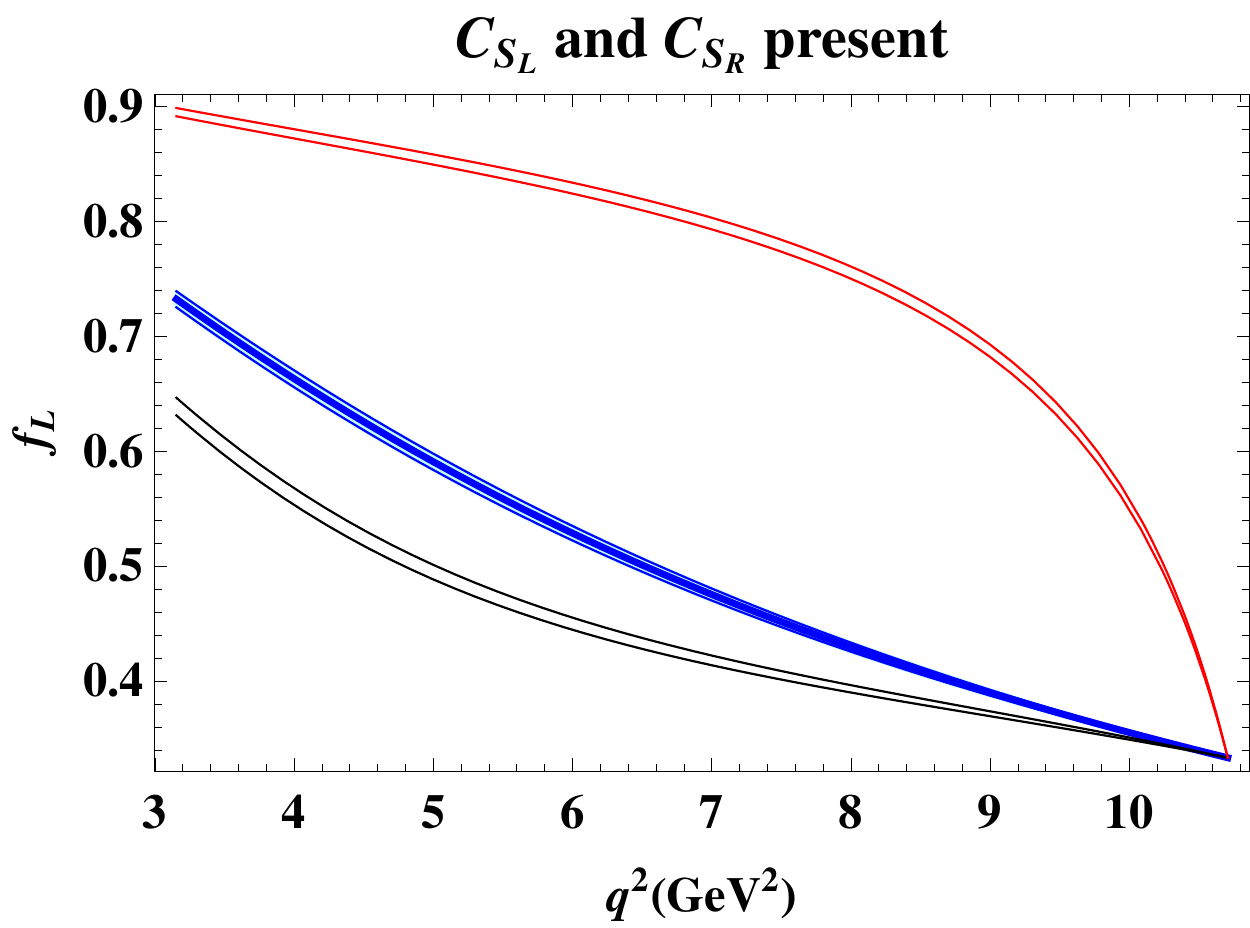}&
\includegraphics[width=35mm]{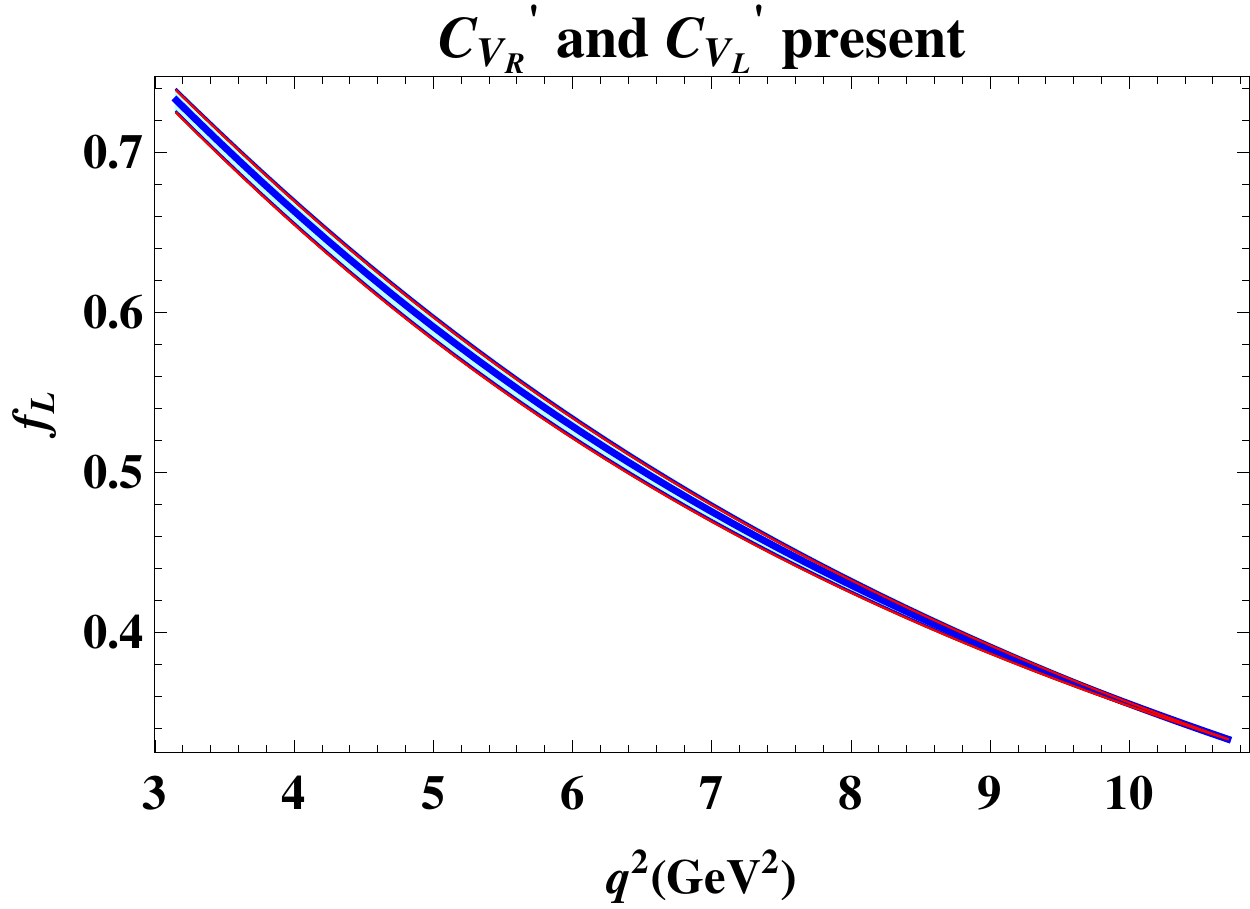}&
\includegraphics[width=35mm]{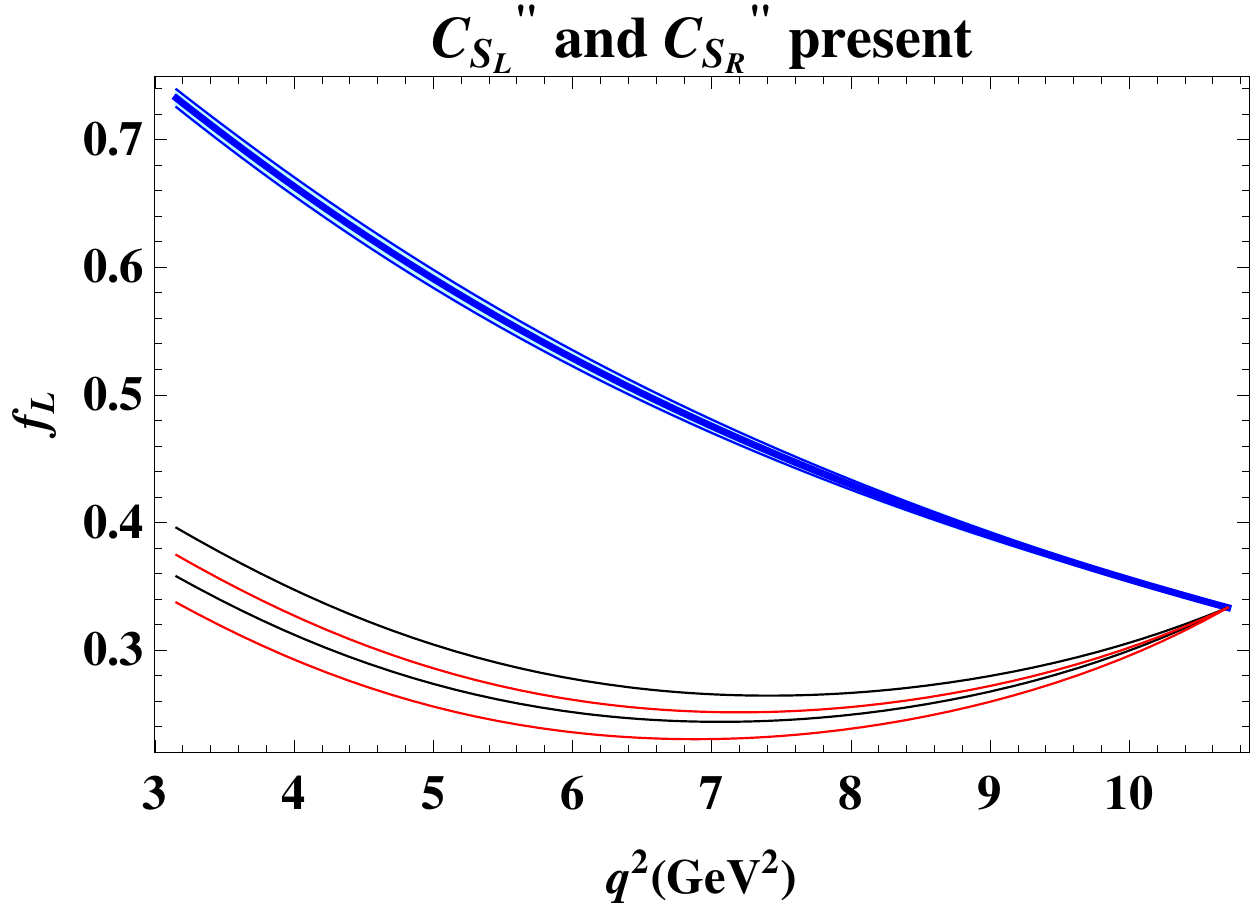}\\
\end{tabular}
\caption{Plots of $D^*$ longitudinal polarization fraction $f_L(q^2)$ as a function of  the dilepton invariant mass $q^2$ in the decay $\bar{B} \to D^* \tau \bar{\nu}$. The blue band in all the plots corresponds to the SM prediction. The band is due to theoretical uncertainties, mainly due to form factors, added in quadrature. The plot in the left panel of the top row represents 
$f_L(q^2)$ prediction in the presence of NP couplings $C_{V_L} = (0.18\pm 0.04)$ (black band) and $C_{V_L} = (-2.88\pm 0.04)$ (red band). The black and red bands in the middle panel of the top
row are for $f_{L}(q^2)$ with NP couplings $C_{T} = (0.52\pm 0.02)$ and  $C_{T} = 
(-0.07\pm 0.02)$, respectively. 
The red band in the right panel of top row corresponds to $C^{''}_{S_L} = (-0.46\pm 0.09)$. 
In the left panel of the bottom row, the black and red bands correspond to NP coefficients $(C_{S_L},C_{S_R}) = (-1.02,1.25)$ and $ (3.08,-2.84)$, respectively.  
In the bottom middle panel, $f_{L}(q^2)$ prediction for  
$(C^{'}_{V_L},C^{'}_{V_R}) = (0.18,-0.01)$ and  $ (-2.88,0.01)$ are shown  by black and red bands, respectively. $f_{L}(q^2)$ for NP couplings $(C^{''}_{S_R},C^{''}_{S_L}) = (0.96,2.41)$  (black band) and  $ (-6.34,-2.39)$ (red band) are shown in right panel of bottom row. (Color online)}
\label{flq-mi}
\end{figure*}

Fig.~\ref{flq-mi} depicts $f_L(q^2)$ in different panels for different NP operators given in Table~\ref{operator}. In these panels, the blue curve represents the SM and the red and the black curves represent NP operators. Each curve is in the form of a narrow band. The thickness of the band represents the theoretical uncertainty in $f_L(q^2)$, mainly due to form factor \cite{Tanaka:2012nw}, which is quite small. We discuss the form of $f_L(q^2)$ panel by panel: 
\begin{itemize}
\item \underline{Only $C_{V_L}$ present:} This NP has the same Lorentz structure of the SM operator. Hence $f_L(q^2)$ for this case has a complete overlap with SM prediction. 
\item \underline{Only $C_{T}$ present:} Here there are two solutions, one with a large value of $C_T$ and the other with a small value. It is difficult to distinguish the small $C_T$ case from SM but $f_L(q^2)$, for the large $C_T$ case, is much smaller than SM prediction for the whole range of $q^2$. Therefore $\langle f_L(q^2) \rangle$ is quite distinguishable from the SM prediction.
\item \underline{Only $C''_{S_L}$ present:} Here the Lorentz structure is different from that of SM. But $f_L(q^2)$ is nearly the same as that of SM because the coupling constant is quite small.
\item \underline{$C_{S_L}$ and $C_{S_R}$ present:} These Lorentz structures are quite different from SM and NP couplings are moderately large for both the allowed solutions. Hence $f_L(q^2)$ for both of them is significantly different from SM. $\langle f_L (q^2) \rangle$ can distinguish 
$(-2.84,3.08)$ solution from the SM but not $(1.25,-1.02)$ solution. To achieve such a distinction, an accurate measurement of $f_L(q^2)$ at low $q^2$ is needed.  
\item \underline{$C'_{V_R}$ and $C'_{V_L}$ present:} For both the allowed solutions $C'_{V_R}$ is negligibly small. Therefore the NP has the same Lorentz structure of SM and hence $f_{L}(q^2)$ cannot distinguish it from SM.
\item \underline{$C''_{S_R}$ and $C''_{S_L}$ present:} For the two solution allowed by \cite{Li:2016vvp} $C''_{S_L}$ is negligibly small, leaving a significant coefficient only for $\mathcal{O}''_{S_R}$ which has the same Lorentz structure of SM. Hence $f_L(q^2)$ cannot distinguish these two solutions from SM. The two disallowed solutions have large $C''_{S_L}$ values and 
 the values  $f_L(q^2)$ and $\langle f_L(q^2) \rangle$ for these are significantly different from the SM because $\mathcal{O}''_{S_L}$ has a different Lorentz structure from SM.

\end{itemize}

A number of papers tried to account for $R_{D^*}$ anomaly through leptoquark models, see for e.g., \cite{Sakaki:2013bfa,Fajfer:2015ycq,Bauer:2015knc,Sahoo:2016pet}. We find that $f_L(q^2)$ cannot discriminate amongst any of these models. This is because their Fierz transformed operators have the Lorentz structure either $\mathcal{O}_{V_L}$ or $\mathcal{O}_{V_R}$.  However in some of the leptoquark models, such as  those discussed in \cite{Dorsner:2013tla,Dorsner:2016wpm,Chen:2017hir}, the $b \to c \tau \nu$ transitions occur through either $\mathcal{O}'_{S_L}$ or $\mathcal{O}''_{S_L}$ operators, whose Fierz transforms are linear combinations of $\mathcal{O}_{S_L}$ and $\mathcal{O}_{T}$. In such cases, the $D^*$ polarization can lead to a discrimination provided the couplings of $\mathcal{O}'_{S_L}$/$\mathcal{O}''_{S_L}$ operators are large enough.

\begin{figure*}[h!] 
\centering
\begin{tabular}{ccc}
\includegraphics[width=35mm]{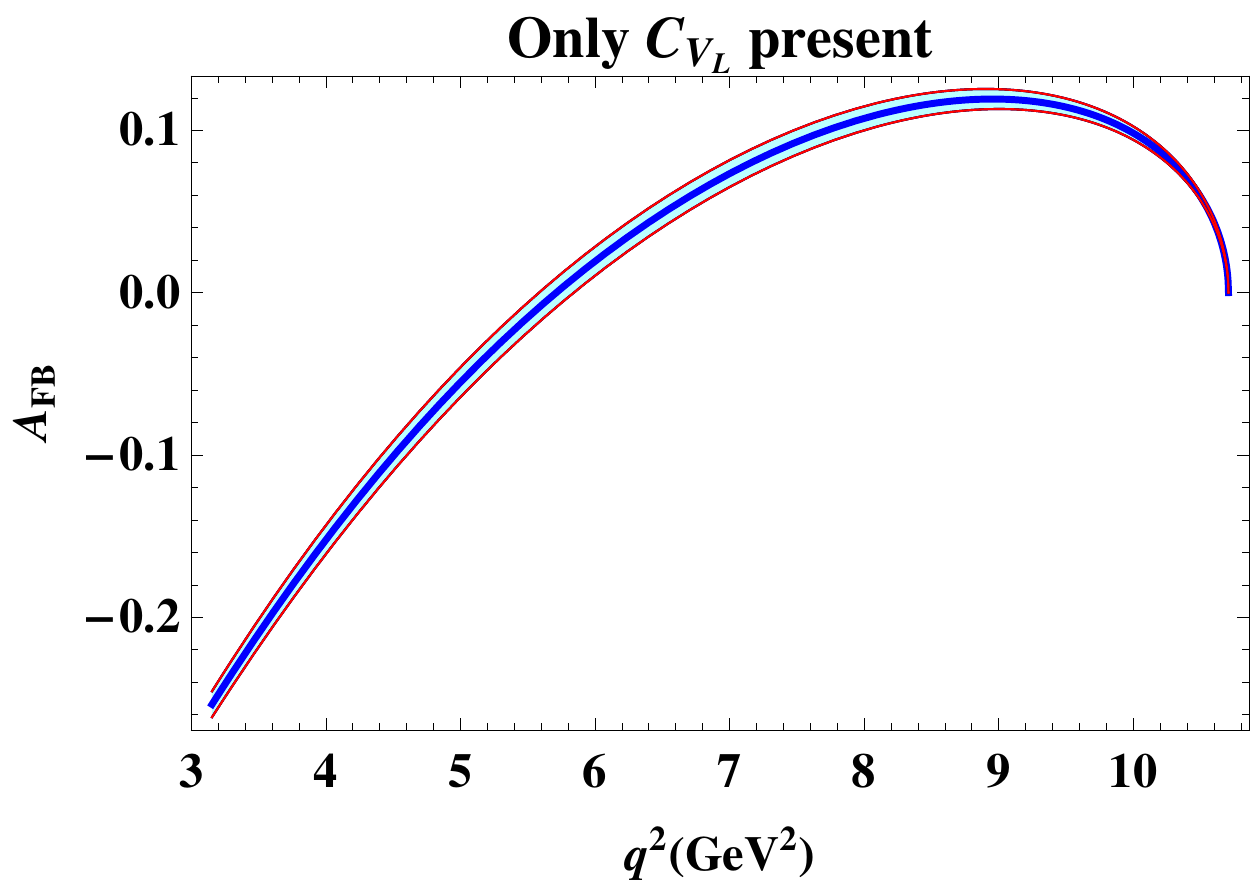}&
\includegraphics[width=35mm]{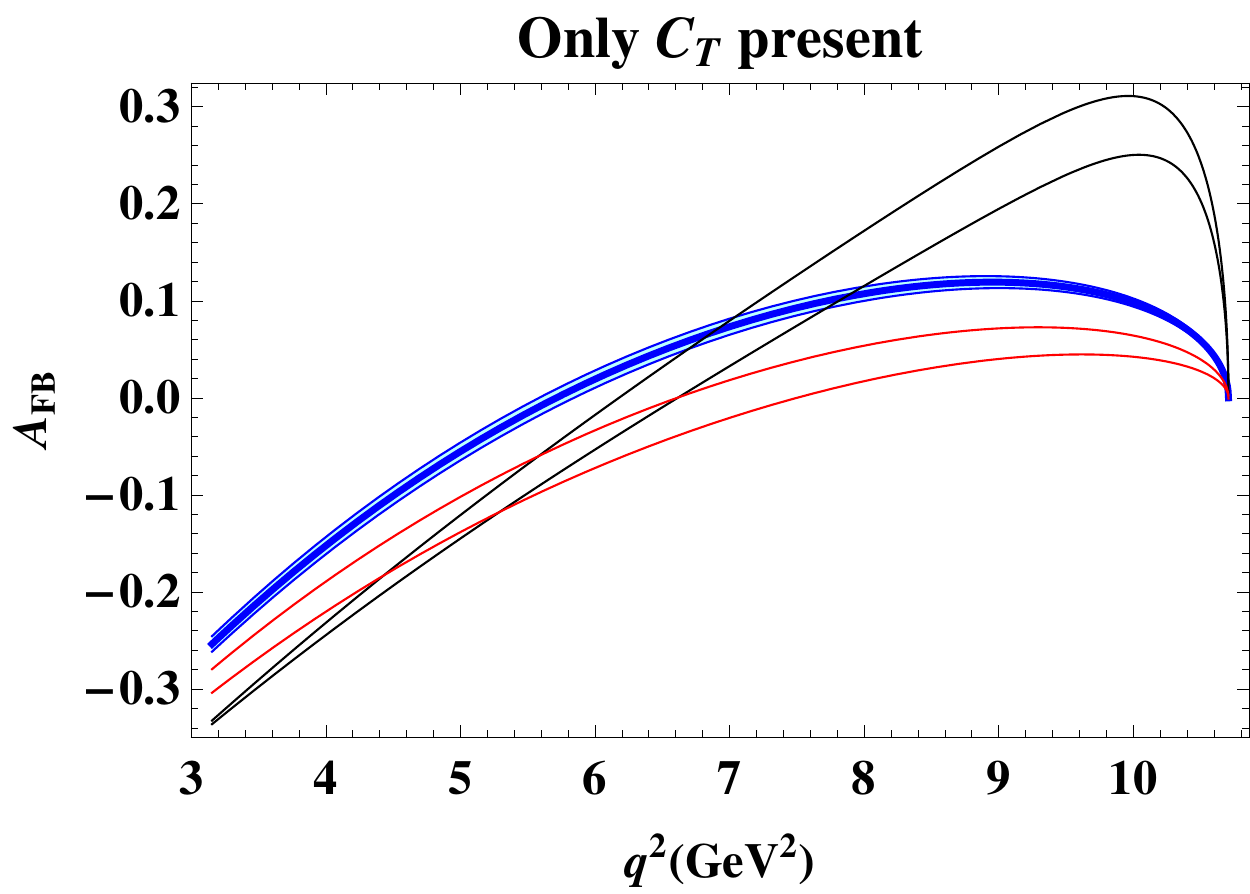}&
\includegraphics[width=35mm]{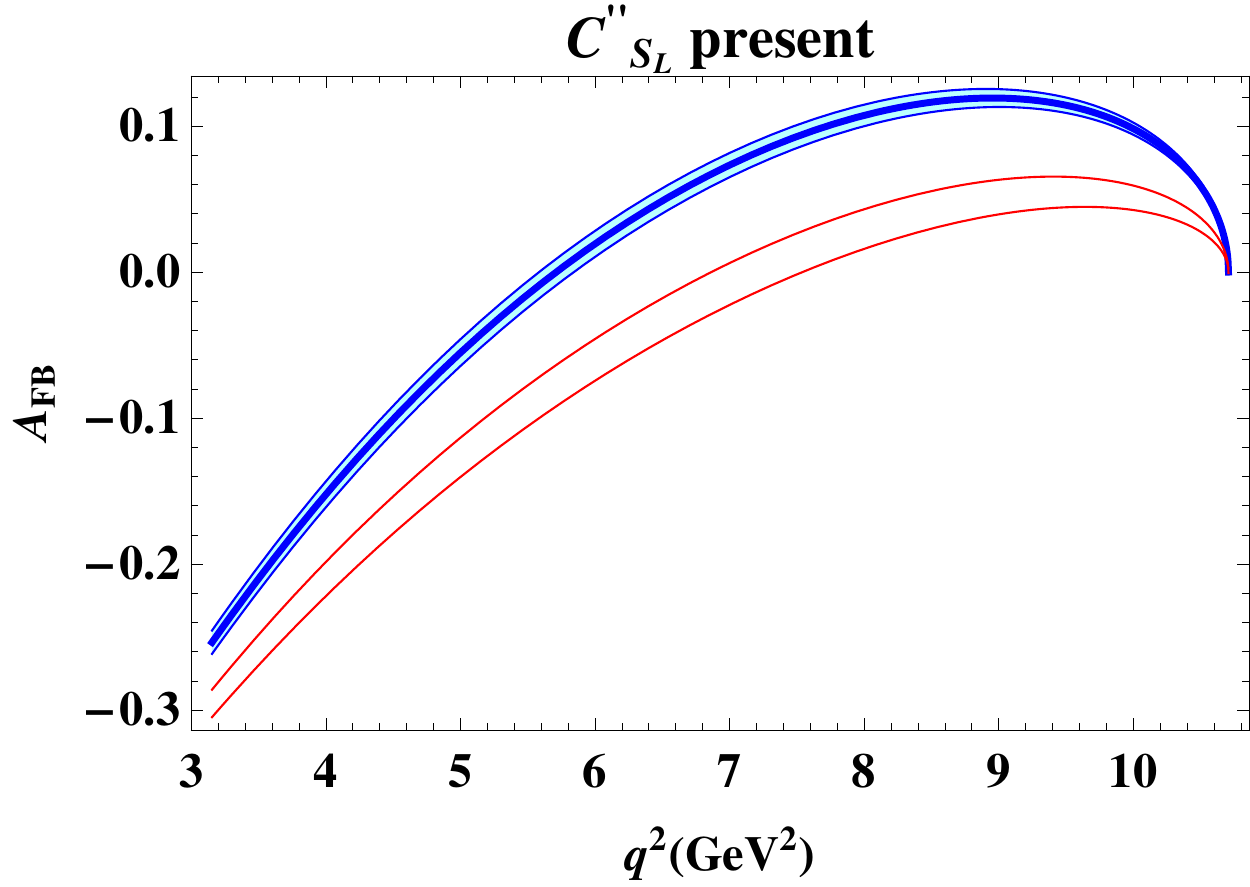}\\
\includegraphics[width=35mm]{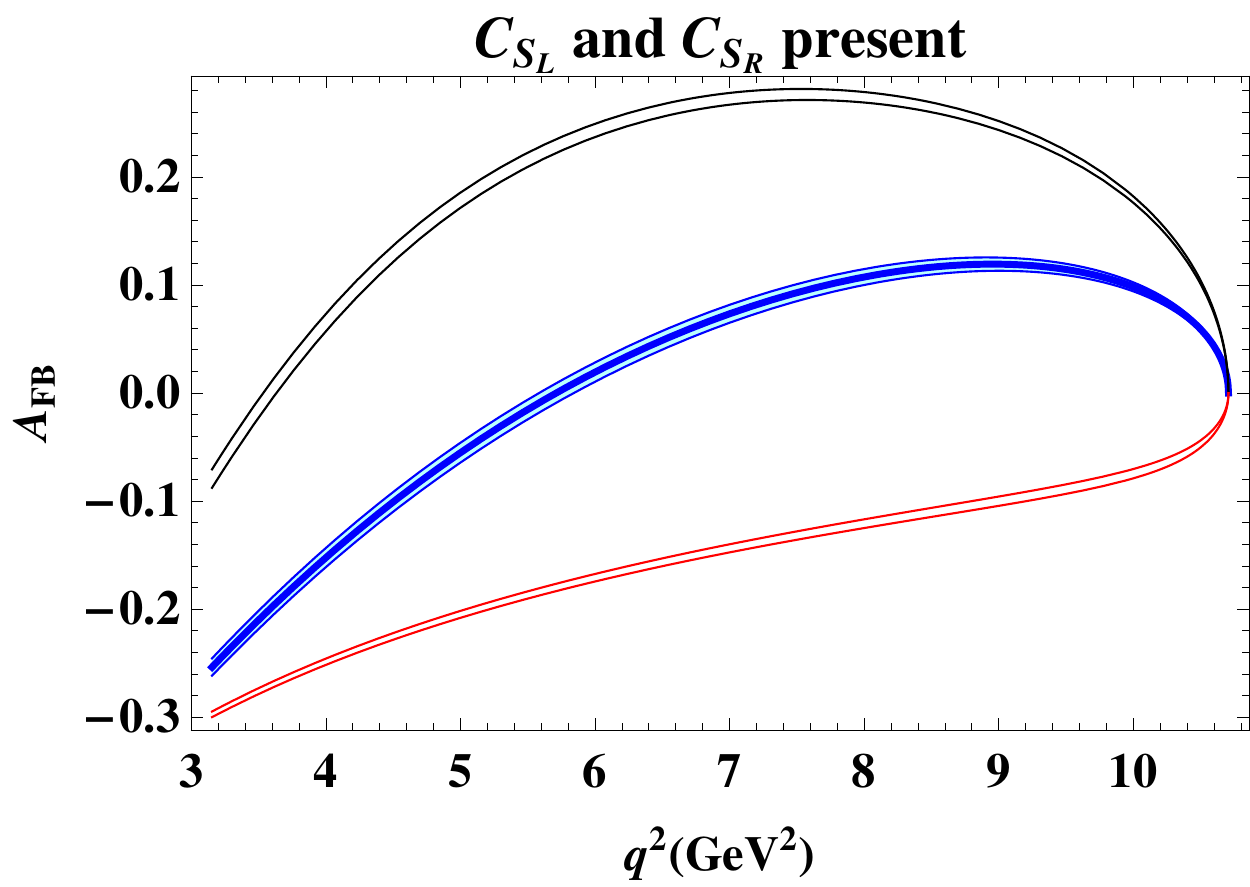}&
\includegraphics[width=35mm]{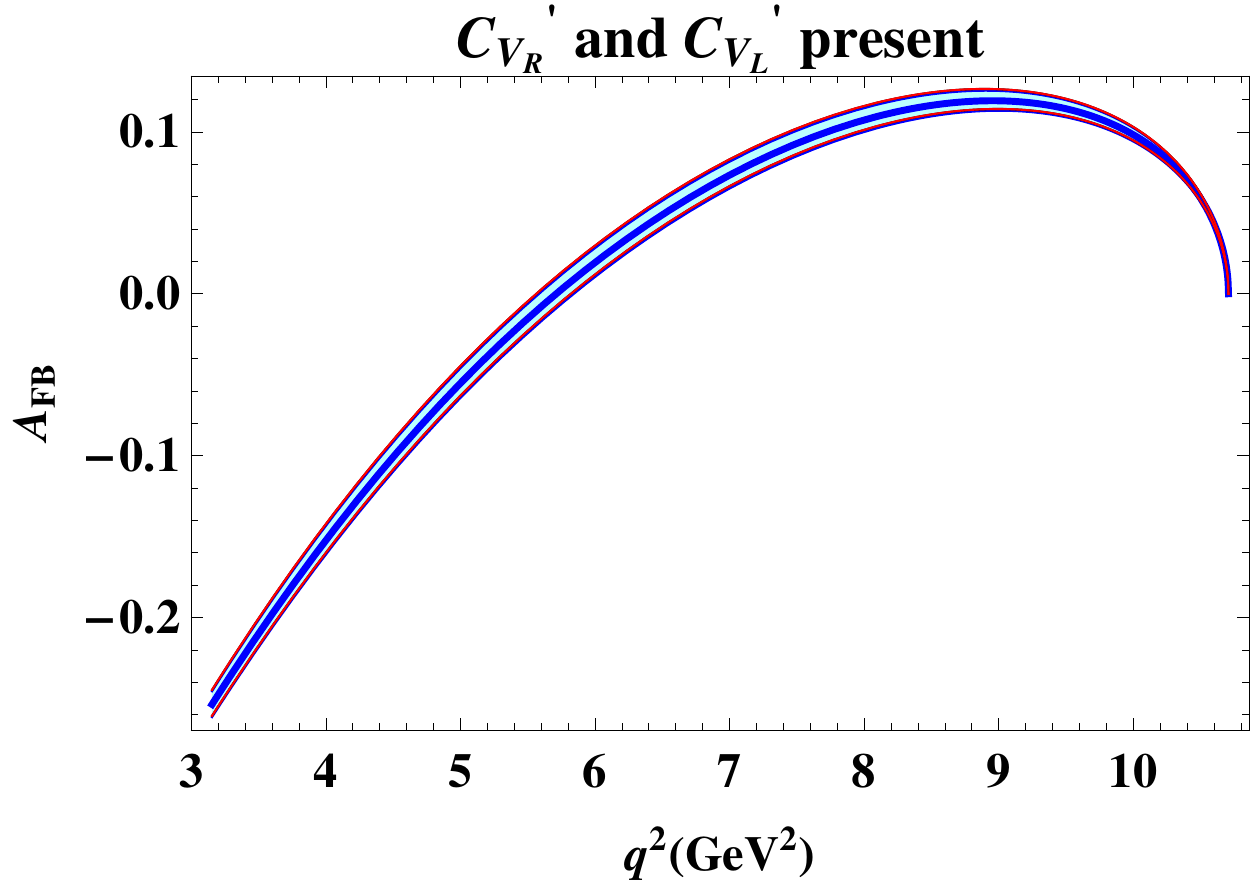}&
\includegraphics[width=35mm]{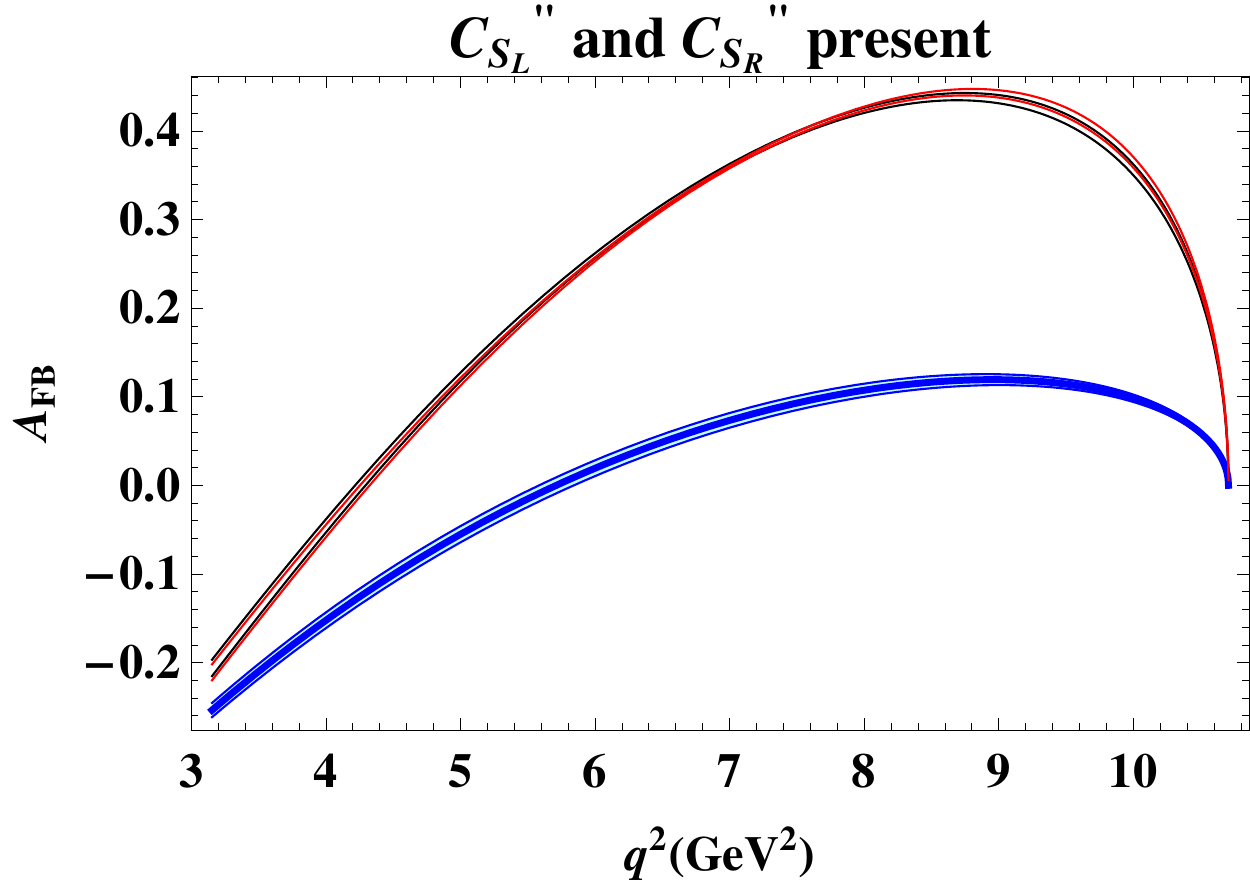}\\
\end{tabular}
\caption{Plots of lepton forward-backward asymmetry $A_{FB}(q^2)$ as a function of  the dilepton invariant mass $q^2$ in the decay $\bar{B} \to D^* \tau \bar{\nu}$. The blue band in all the plots corresponds to the SM prediction. The band is due to theoretical uncertainties, mainly due to form factors, added in quadrature. The plot in the left panel of top row represents  $A_{FB}(q^2)$ prediction in the presence of NP couplings $C_{V_L} = (0.18\pm 0.04)$ (black band) and $C_{V_L} = (-2.88\pm 0.04)$ (red band). The black and red bands in the middle panel of the top row represent NP couplings $C_{T} = (0.52\pm 0.02)$ and  $C_{T} = (-0.07\pm 0.02)$, respectively. The red band in the right panel of top row corresponds to $C^{''}_{S_L} = (-0.46\pm 0.09)$. In the left panel of bottom row, the black and red bands correspond to $A_{FB} (q^2)$ with NP coefficients $(C_{S_L},C_{S_R}) = (-1.02,1.25)$ and $ (3.08,-2.84)$, respectively.  In the bottom middle panel, $A_{FB}(q^2)$ prediction for  $(C^{'}_{V_L},C^{'}_{V_R}) = (0.18,-0.01)$ and  $ (-2.88,0.01)$ are shown  by black and red bands, respectively. $A_{FB}(q^2)$ with NP couplings $(C^{''}_{S_R},C^{''}_{S_L}) = (0.96,2.41)$  (black band) and  $ (-6.34,-2.39)$ (red band) are shown in right panel of bottom row. (Color online)}
\label{afb}
\end{figure*}

It would be interesting to see the discrimination capability of the asymmetries based on other angular observables. One such example is the forward-backward asymmetry of leptons, $A_{FB}(q^2)$ in $\bar{B} \to D^* \tau \bar{\nu}$. The values of $A_{FB}(q^2)$ for the six different NP combinations listed in Table~\ref{tab1} are plotted in various panels of fig.~\ref{afb}. Here again the blue band represents the SM and the red and the black bands represent NP solutions. As in the case of $f_L(q^2)$, the plots of $A_{FB}(q^2)$ distinguish the NP solutions from SM for the following cases:(i) Only $C_{T}$ present, (ii) $C_{S_L}$ and $C_{S_R}$ present and (iii) the two disallowed solutions of $C''_{S_R}$ and $C''_{S_L}$ present. For the other three cases,
$A_{FB}(q^2)$ is either same or differs very little from the SM values. 
From figs.~\ref{flq-mi} and~\ref{afb}, we note that $f_L(q^2)$ has better
discrimination capability for the $C_{T} = (0.52\pm 0.02)$ solution compared to $A_{FB}(q^2)$,
whereas the situation is reverse for the $C^{''}_{S_L} = (-0.46\pm 0.09)$ solution.  

 Determination of $A_{FB}(q^2)$ requires the reconstruction of $\tau$ momentum, which is difficult because of the missing neutrino/neutrinos in the final state. It may be possible for LHCb to reconstruct $\tau$ in those events where $\tau$ decays into multiple hadrons, using the same technique they used to identify the $B$ meson in the decay $B \rightarrow D^* \tau \, \bar{\nu}$. But as demonstrated in figs. \ref{flq-mi} and \ref{afb}, such a measurement leads
only to a small advantage in distinguishing between allowed NP models.

The Belle collaboration is in the process of measuring $\langle f_L(q^2) \rangle$. It is expected that the uncertainty in this measurement will be
about $\pm 0.1$ \cite{belle-ckm16}. The SM prediction for this quantity is $0.46 \pm 0.04$, where the uncertainity comes from the erros in the form-factors, the details of which are given in Appendix. As we can see from Table~\ref{tab1}, $\langle f_L(q^2) \rangle = 0.14 \pm 0.04$ for NP coupling $C_T=0.52$ and  $\langle f_L(q^2) \rangle = 0.76 \pm 0.04$ for $(C_{S_R}, \,C_{S_L})=(-2.84,\, 3.08)$. For these two cases, the change in $\langle f_L(q^2) \rangle$ is {\it three} times the expected uncertainty in its measurement. Therefore the upcoming Belle measurement can confirm one of these two NP solutions or rule them out at better than 95 \% C.L. The $\langle f_L(q^2) \rangle$ can also discriminate two other solutions with $(O^{''}_{S_R}, \,O^{''}_{S_L})$ operators but, as shown in \cite{Li:2016vvp}, these are already  ruled out by their prediction of $B_c \to \tau\, \nu$ partial width.

 As seen from Figs.~\ref{flq-mi} and \ref{afb}, angular asymmetries $f_L$ and $A_{FB}$ are not sensitive to the presence of right handed currents. This is because $B \to D$ transition occurs purely through vector current and $B \to D^*$ transition occurs purely through axial-vector currents. However $\tau$ polarization will be a good discriminant of right handed currents. The sensitivity of $\tau$ polarization to new physics in  $B \to D^*\, \tau\,\nu$ is discussed in \cite{Alonso:2016oyd}. For the decay $B \to D\, \tau\,\nu$, the corresponding discussion is given in  \cite{Alonso:2017ktd}.

Recently Belle Collaboration \cite{Abdesselam:2016xqt} 
used a new technique
to identify the $\tau$ lepton in the decay $\bar{B} \to D^* \tau \nu_\tau$,
through the decays $\tau \to \pi \nu_\tau$ and $\tau \to \rho \nu_\tau$.
This leads to a reduced signal size. With such a signal
definition, they obtained $R_{D^*} = 0.276 \pm 0.034 \pm 0.026$.
This measurement differs from the Standard Model prediction by only 
$0.6 \sigma$, though one must note that the error in this measurement
is twice the error in the current world average. 
A new world average, including this measurement,  is smaller by 3\% compared to the older value. Hence, we believe that our results will not change much by the inclusion of this new result.
Therefore it is worthwhile to develop signatures which can  help in discerning effects of new physics
in this decay.

\section{Conclusions}
In this work, we studied the possibility to distinguish between NP solutions which can explain the observed excess in $R_{D^*}$. The angular observables in the decay $B \to D^* \tau \bar{\nu}$ can discriminate between some of the scalar and tensor NP solutions. The $D^*$ polarization, $f_L(q^2)$, and the lepton forward-backward asymmetry, $A_{FB}(q^2)$, are both capable of this discrimination. A measurement of $A_{FB}(q^2)$ is more difficult as it requires the $\tau$ reconstruction. Belle collaboration is in the process of measuring $\langle f_L (q^2) \rangle$. Such a measurement can confirm or rule out two of the NP solutions at better than 95 \% C.L. 
\label{concl}

{\em Acknowledgments.\textemdash}
We thank Karol Adamczyk for numerous discussions regarding the measurement of  $f_L(q^2)$ at Belle.

\begin{appendix}
\section{${B\to \Dst}$ from factors}

The $B \to D^*\tau\nu$ vector and axial vector operator matrix elements, which depend on the momentum transfer between $B$ and $D^*$ , can be expressed as
\begin{equation}
   \begin{split}
      \langle\Dst(k,\varepsilon)|\cbar\gamma_\mu b|\Bbar(p)\rangle =& -i\epsilon_{\mu\nu\rho\sigma}\varepsilon^{\nu*}p^\rho k^\sigma{2V(q^2) \over m_B+m_\Dst} \,, \\
      \langle\Dst(k,\varepsilon)|\cbar\gamma_\mu\gamma_5 b|\Bbar(p)\rangle =& \varepsilon^{\mu*}(m_B+m_\Dst)A_1(q^2) - (p+k)_\mu(\varepsilon^*q){A_2(q^2) \over m_B+m_\Dst} \\
      & -q_\mu(\varepsilon^*q){2m_\Dst \over q^2}[A_3(q^2)-A_0(q^2)] \,,
   \end{split}
   \label{eq:VA_parametrization}
\end{equation}
\begin{equation}
   \begin{split}
      \langle\Dst(k,\varepsilon)|\cbar\gamma_5 b|\Bbar(p)\rangle =& -{1 \over m_b+m_c}q_\mu\langle\Dst(k,\varepsilon)|\cbar\gamma^\mu\gamma^5 b|\Bbar(p)\rangle \\
      =& -( \varepsilon^* q ) { 2m_\Dst \over m_b + m_c }A_0(q^2) \,.
   \end{split}
   \label{eq:scalar_parametrization_Dst}
\end{equation}

\begin{equation}
   \begin{split}
      \langle\Dst(k,\varepsilon)|\cbar\sigma_{\mu\nu} b|\Bbar(p) & \rangle = \epsilon_{\mu\nu\rho\sigma} \biggl\{ -\varepsilon^{*\rho} (p+k)^\sigma T_1(q^2) \biggr. \\
                                                                 & + \varepsilon^{*\rho} q^\sigma { m_B^2 - m_\Dst^2 \over q^2 } [ T_1(q^2) - T_2(q^2) ] \\
                                                                 & \biggl. + 2 { (\varepsilon^* \cdot q) \over q^2 } p^\rho k^\sigma \left[ T_1(q^2) - T_2(q^2) - { q^2 \over m_B^2 - m_\Dst^2 } T_3(q^2) \right] \biggr\} \,,
   \end{split}
   \label{eq:T_parametrization}
\end{equation}
\begin{equation}
   \begin{split}
      \langle\Dst(k,\varepsilon)|\cbar\sigma_{\mu\nu}q^\nu b|\Bbar(p)\rangle =& \epsilon_{\mu\nu\rho\sigma}\varepsilon^{*\nu}p^\rho k^\sigma2T_1(q^2) \,, \\
      \langle\Dst(k,\varepsilon)|\cbar\sigma_{\mu\nu}\gamma_5 q^\nu b|\Bbar(p)\rangle =& -\left[(m_B^2-m_\Dst^2)\varepsilon^{*\mu} - (\varepsilon^* q)(p+k)_\mu\right]T_2(q^2) \\
      & - (\varepsilon^* q)\left[q_\mu-{q^2 \over m_B^2-m_\Dst^2}(p+k)_\mu\right]T_3(q^2) \,.
   \end{split}
   \label{eq:T_parametrization_2}
\end{equation}

where
\begin{equation}
   A_3(q^2) = {m_B+m_\Dst \over 2m_\Dst}A_1(q^2)-{m_B-m_\Dst \over 2m_\Dst}A_2(q^2) \,,
\end{equation}
with $A_3(0)=A_0(0)$.
The form factors $V,A_0,A_1,A_2,T_1,T_2,T_3$ can be written in terms of the heavy quark effective theory (HQET) form factors as \cite{Sakaki:2013bfa,Caprini:1997mu}
\begin{equation}
   \begin{split}
      V(q^2) =& { m_B + m_\Dst \over 2\sqrt{m_B m_\Dst} } \, h_V(w(q^2)) \,, \\
      A_1(q^2) =& { ( m_B + m_\Dst )^2 - q^2 \over 2\sqrt{m_B m_\Dst} ( m_B + m_\Dst ) } \, h_{A_1}(w(q^2)) \,, \\
      A_2(q^2) =& { m_B+m_\Dst \over 2\sqrt{m_B m_\Dst} } \left[ h_{A_3}(w(q^2)) + { m_\Dst \over m_B } h_{A_2}(w(q^2)) \right] \,, \\
      A_0(q^2) =& { 1 \over 2\sqrt{m_B m_\Dst} } \left[ { ( m_B + m_\Dst )^2 - q^2 \over 2m_\Dst } \, h_{A_1}(w(q^2)) \right. \\
                & -\left. { m_B^2 - m_\Dst^2 + q^2 \over 2m_B } \, h_{A_2}(w(q^2)) - { m_B^2 - m_\Dst^2 - q^2 \over 2m_\Dst } \, h_{A_3}(w(q^2)) \right] \,,
   \end{split}
   \label{eq:VA-hVA_relation}
\end{equation}

\begin{equation}
   \begin{split}
      T_1(q^2) =& { 1 \over 2\sqrt{m_B m_\Dst} } \left[ ( m_B + m_\Dst ) h_{T_1}(w(q^2)) - ( m_B - m_\Dst ) h_{T_2}(w(q^2)) \right] \,, \\
      T_2(q^2) =& { 1 \over 2\sqrt{m_B m_\Dst} } \left[ { ( m_B + m_\Dst )^2 - q^2 \over m_B + m_\Dst } \, h_{T_1}(w(q^2)) \right. \\
                & \quad\quad\quad\quad\quad\quad \left. - { ( m_B - m_\Dst )^2 - q^2 \over m_B - m_\Dst } \, h_{T_2}(w(q^2)) \right] \,, \\
      T_3(q^2) =&  { 1 \over 2\sqrt{m_B m_\Dst} } \left[ ( m_B - m_\Dst ) h_{T_1}(w(q^2)) - ( m_B + m_\Dst ) h_{T_2}(w(q^2)) \right. \\
                & \quad\quad\quad\quad\quad\quad \left.- 2 { m_B^2 -m_\Dst^2 \over m_B } h_{T_3}(w(q^2)) \right] \,.
   \end{split}
   \label{eq:T-hT_relation}
\end{equation}
where the HQET form factors can be expressed as \cite{Caprini:1997mu}
   \begin{align}
      \begin{split}
         h_V(w) =& R_1(w) h_{A_1}(w) \,, \\
         h_{A_2}(w) =& { R_2(w)-R_3(w) \over 2\,r_\Dst } h_{A_1}(w) \,, \\
         h_{A_3}(w) =& { R_2(w)+R_3(w) \over 2 } h_{A_1}(w) \,,\\
       h_{T_1}(w) =& { 1 \over 2 ( 1 + r_\Dst^2 - 2r_\Dst w ) } \left[ { m_b - m_c \over m_B - m_\Dst } ( 1 - r_\Dst )^2 ( w + 1 ) \, h_{A_1}(w) \right. \\
                     & \quad\quad\quad\quad\quad\quad\quad\quad\quad \left. - { m_b + m_c \over m_B + m_\Dst } ( 1 + r_\Dst )^2 ( w - 1 ) \, h_V(w) \right] \,, \\
         h_{T_2}(w) =& { ( 1 - r_\Dst^2 ) ( w + 1 ) \over 2 ( 1 + r_\Dst^2 - 2r_\Dst w ) } \left[ { m_b - m_c \over m_B - m_\Dst } \, h_{A_1}(w) - { m_b + m_c \over m_B + m_\Dst } \, h_V(w) \right] \,, \\
         h_{T_3}(w) =& -{ 1 \over 2 ( 1 + r_\Dst ) ( 1 + r_\Dst^2 - 2r_\Dst w ) } \left[2 { m_b - m_c \over m_B - m_\Dst } r_\Dst ( w + 1 ) \, h_{A_1}(w) \right. \\
                     & - { m_b - m_c \over m_B - m_\Dst } ( 1 + r_\Dst^2 - 2r_\Dst w ) ( h_{A_3}(w) - r_\Dst h_{A_2}(w) )`` \\
                     & \left. - { m_b + m_c \over m_B + m_\Dst } ( 1 + r_\Dst )^2 \, h_V(w) \right] \,,
      \end{split}
   \end{align}
where the $w$-dependencies are parametrized as \cite{Caprini:1997mu}
\begin{equation}
   \begin{split}
      h_{A_1}(w) =& h_{A_1}(1) [ 1 - 8\rho_\Dst^2 z + (53\rho_\Dst^2-15) z^2 - (231\rho_\Dst^2-91) z^3 ] \,, \\
      R_1(w) =& R_1(1) - 0.12(w-1) + 0.05(w-1)^2 \,, \\
      R_2(w) =& R_2(1) + 0.11(w-1) - 0.06(w-1)^2 \,, \\
      R_3(w) =& 1.22 - 0.052(w-1) + 0.026(w-1)^2 \,,
      \label{eq:HQET_parametrization}
   \end{split}
\end{equation}
where $r_{D^{*}} = M_{D^{*}}/M_B$\,,$w = (M_B^2+M_{D^*}^2-q^2)/2M_BM_{D^*}$\,, $z(w) = ( \sqrt{w+1} - \sqrt2 ) / ( \sqrt{w+1} + \sqrt2 )$.

The numerical values of some of the parameters used in form factors are given by
\begin{align}
       h_{A_1}(1) =& 0.908 \pm 0.017\,\,\text{\cite{Bailey:2014tva}},&&\quad~~\rho_\Dst^2 = 1.207 \pm 0.026\,\,\text{\cite{Amhis:2014hma}}, \\
      R_1(1) =& 1.403 \pm 0.033\,\,\text{\cite{Amhis:2014hma}},&& \quad R_2(1) = 0.854 \pm 0.020\,\,\text{\cite{Amhis:2014hma}}\,.
   \label{eq:HQET_fit}
\end{align}

\end{appendix}

\end{document}